\documentclass[
prd %, twocolumn%
,showpacs, amssymb, superscriptaddress, aps, dvipdfmx
]{revtex4}
\usepackage{graphicx}
\input{epsf}

\usepackage{amsmath,amssymb}
\usepackage{bm}
\usepackage{times}

\newcommand{\dalm}{\kern1pt\vbox{\hrule height 0.9pt\hbox{\vrule width 0.9pt
\hskip 2.5pt\vbox{\vskip 5.5pt}\hskip 3pt\vrule width 0.3pt}\hrule height 0.3pt}
\kern1pt}

\newcommand{\gsim}{\, \raisebox{-0.8ex}{$\stackrel{\textstyle >}{\sim}$ }}

\newcommand{\be}{\begin{eqnarray}}
\newcommand{\ee}{\end{eqnarray}}
\newcommand{\beq}{\begin{eqnarray}}
\newcommand{\eeq}{\end{eqnarray}}

\begin{document}

%\twocolumn[\hsize\textwidth\columnwidth\hsize\csname @twocolumnfals\endcsname

% For two column
%\wideabs{

\title{Light curves from highly compact neutron stars with spot size effect}

\author{Hajime Sotani}
%\author{Hajime Sotani\,\orcidlink{0000-0002-3239-2921}}
\email{sotani@yukawa.kyoto-u.ac.jp}
\affiliation{Astrophysical Big Bang Laboratory, RIKEN, Saitama 351-0198, Japan}
\affiliation{Interdisciplinary Theoretical \& Mathematical Science Program (iTHEMS), RIKEN, Saitama 351-0198, Japan}

%\author{Hector O. Silva}
%\email{sotani@yukawa.kyoto-u.ac.jp}
%\affiliation{eXtreme Gravity Institute, Department of Physics, Montana State University, Bozeman, Montana 59717 USA}

%\author{George Pappas}
%\email{sotani@yukawa.kyoto-u.ac.jp}
%\affiliation{Dipartimento di Fisica, ``Sapienza" Universit\'{a} di Roma \& Sezione INFN Roma1, Piazzale Aldo Moro 5, 00185, Roma, Italy}

\date{\today}

% Abstract
\begin{abstract}
We systematically examine light curves from a single hot spot on a slowly rotating neutron star with very high compactness, where the  so-called invisible zone does not exist. In particular, we adopt three different shapes of hot spot and  take into account the finite size effect of hot spot on the light curves. Then, we find that the brightening of flux occurs when the hot spot with small area crosses the opposite side to the observer, where the brightening becomes stronger as the area of hot spot decreases. Since this brightening happens only when a part of hot spot crosses the opposite side to the observer, one may constrain the neutron star geometry, i.e., the combination of an inclination angle, angle between the rotational axis and normal vector at the center of hot spot, and the opening angle of hot spot, if the brightening would be observed. In addition, by counting such a brightening in the light curves, one may know how many narrow bands of hot spot and/or spots with small area cross the opposite side to the observer.  
\end{abstract}

%\pacs{04.80.Cc, 95.30.Sf}
%
%%%%%%%%%%%%%%%%%%%%%%%%%%%%%%%%%%%%%%%%%%%%%%%%%
%  04.40.Dg :  Relativistic stars: structure, stability, and oscillations (see also 97.60.-s Late stages of stellar evolution) 
%  04.40.Nr  :  Einstein-Maxwell spacetimes, spacetimes with fluids, radiation or classical fields
%  04.50.Kd : Modified theories of gravity
%  04.70.-s   : Physics of black holes
%  04.80.Cc : Experimental tests of gravitational theories
%  95.30.Sf : Relativity and gravitation (95.30.-k : Fundamental aspects of astrophysics)
%  26.60.Kp : Equations of state of neutron-star matter
%%%%%%%%%%%%%%%%%%%%%%%%%%%%%%%%%%%%%%%%%%%%%%%%%
%]
% For two column
%}

\maketitle

%\baselineskip 24pt
%%%%%%%%%%%%%%%%%%%%%%%%%%%%%%%%%%%%%%%%%%%%%%%%
\section{Introduction}
\label{sec:I}
%%%%%%%%%%%%%%%%%%%%%%%%%%%%%%%%%%%%%%%%%%%%%%%%

Neutron stars, which are produced via supernova explosions at the last moment of massive stars, are a unique object for probing the physics under extreme conditions. The density inside the star significantly exceeds the standard nuclear density, while the gravitational  and magnetic fields can become quite stronger than those observed in our solar system \cite{shapiro-teukolsky}. Due to such extreme conditions, most of the neutron star properties are still uncertain. But then, one could extract an aspect of the physics under these conditions via the observations of neutron stars itself and/or the phenomena associated with neutron stars as an inverse problem. In fact, the discovery of the $2M_\odot$ neutron stars excluded some of the soft equation of state (EOS) for neutron star matter \cite{D2010,A2013,C2019}, and the observations of the quasi-periodic oscillations in the afterglow following the giant flares showed the possibility for constraining the EOS of neutron star crust by identifying the observed frequencies as crustal torsional oscillations (e.g., \cite{SNIO2012,SIO2016}). In addition, the detection of the gravitational wave from the binary neutron stars, GW170817 \cite{GW170817}, enabled us to extract the tidal deformability of neutron stars, which tells us the $1.4M_\odot$ neutron star radius to be less than 13.6 km \cite{AGKV18}. It is also suggested that the future observations of the gravitational waves from compact objects  would not only provide us the stellar properties (e.g., \cite{AK1996,AK1998,STM2001,SH2003,SYMT2011,PA2012,DGKK2013}), but also probe the theory of gravity in strong field regime (e.g., \cite{Berti2015,SK2004,Sotani2014,Sotani2014a}).

As another approach, the observations of the light bending due to the strong gravity around compact objects may also provide us stellar properties. The strong lensing around a black hole is a well-known relativistic effect (e.g., \cite{VE2000,Bozza2002,SM2015}). In the similar way, the light radiated from the neutron star surface (or the light passing in the vicinity of the neutron stars) can be bent due to the strong gravity produced by the neutron stars. In fact, how large the light from the neutron star surface would be bent strongly depends on the neutron star compactness, which is the ratio of the stellar mass to radius. That is, the neutron star with larger compactness can produce a stronger gravitational field, which in turn leads to that the bending angle becomes larger as the stellar compactness increases. As a result, even the photon radiated from the backside of the star may reach the observer, which can change the shape of the light curve from the rotating neutron stars (e.g., \cite{PFC1983,LL95,Beloborodov2002}). Thus, one would see the neutron star compactness via careful observations of the light curves. With this information together with the help of additional observation about the stellar mass or radius, one would constrain the EOS for neutron star matter \cite{POC2014,Bogdanov2016,Watts2016,Watts2019}. We remark that one may also probe the gravitational geometry and the theory of gravity in strong field regime via the observation of the light curve from the compact objects, because the photon trajectory is determined via the gravitational geometry, which depends on the gravitational theory \cite{SM2017,S2017,SY2019a,SY2019b}. Anyway, these observations of light curves will be done in near future with Neutron star Interior Composition ExploreR (NICER) \cite{NICER} and others.

We remark that the stellar compactness for a canonical neutron star is not so high that the photon emitted from a part of the backside of the star can not reach the observer, i.e., an invisible zone exists. The critical value of the neutron star compactness whether or not the invisible zone exits is $M/R=0.2840$ \cite{SM2018a}, above which the invisible zone disappears from the neutron star surface. As shown in Fig. 3 in Ref. \cite{SM2018a}, the region where the invisible zone disappears is relatively narrow in the plot of the relation between the neutron star mass ($M$) and radius ($R$). Namely, the invisible zone exists for most of neutron stars. But, the neutron star with very high compactness where the invisible zone disappears can still exist theoretically and not be excluded observationally. In fact, the compactness of the stellar models with maximum mass constructed with some EOSs, such as APR or SLy, becomes more than $M/R=0.2840$ \cite{example}. If such a highly compact neutron star exists, the expected light curves are significantly different from those for canonical neutron stars \cite{SM2018a}. Thus, this difference, if observed, may become a smoking gun for the existence of highly compact neutron stars. With the same motivation, we have considered the light curves in the previous study, adopting a pointlike approximation, where the area size of hot spot is negligible, but the flux from the side completely opposite to the observer diverges due to the pointike approximation \cite{SM2018a}. So, in this study we take into account the finite size effect of hot spot (as in Refs. \cite{BPO2015,SSP2019}) for the highly compact neutron star and discuss how different the light curves are, depending on the compactness. On the other hand, for considering the light curve from a hot spot on a rotational neutron star, one may have to take into account the rotational effect on the modification of the light curve if the star rotates significantly fast \cite{PG03,PB06,SM2018b,PO2014,CLM05,CMLC07}. In this study, however, we simply consider the light curves from a slowly rotating neutron star, where one can neglect the rotational effects.

In this paper, we adopt geometric units, $c=G=1$, where $c$ and $G$ denote the speed of light and the gravitational constant, respectively, and the metric signature is $(-,+,+,+)$.

%%%%%%%%%%%%%%%%%%%%%%%%%%%%%%%%%%%%%%%%%%%%%%%%
\section{Photon radiating from neutron star surface}
\label{sec:II}
%%%%%%%%%%%%%%%%%%%%%%%%%%%%%%%%%%%%%%%%%%%%%%%%

Since the photon trajectory from a slowly rotating neutron star has been often discussed so far, in this section we briefly mention the basic equations, which are necessary for calculating the photon flux in this study (see for the details in Refs. \cite{SM2018a,SSP2019}). In general, the line element for a static, spherically symmetric spacetime is expressed as
\begin{equation}
   ds^2 = -A(r)dt^2 + B(r)dr^2 + C(r)\left(d\theta^2 + \sin^2 d\psi^2\right), \label{eq:metric}
\end{equation}
where we especially focus on asymptotically flat spacetimes in this study, i.e., $A(r)\to1$, $B(r)\to 1$, and $C(r)\to r^2$ as $r\to \infty$. With this metric form, the circumference radius, $r_c$, is given by $r_c^2\equiv C(r)$ at each radial position, $r$. In this study, we only consider the Schwarzschild spacetime, where the metric functions are given with the neutron star mass, $M$, via
\begin{equation}
   A(r) = 1 - \frac{2M}{r}, \ \ B(r) = \frac{1}{A(r)}, \ \ C(r) = r^2. 
\end{equation}

\begin{figure}[tbp]
\begin{center}
\includegraphics[scale=0.4]{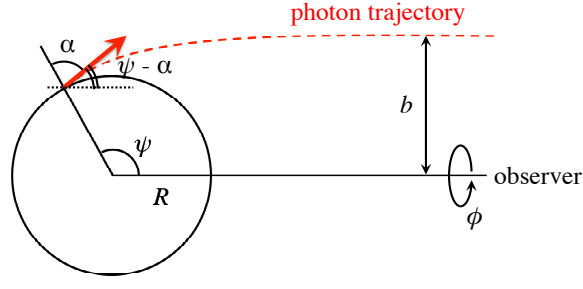}  
\end{center}
\caption{%%
Image of the photon trajectory radiating from the neutron star surface. The photon emits from the angular position $\psi$ with the emission angle $\alpha$, where a distant observer is located at $\psi=0$. The angle $\phi$ is the azimuthal angle in the observer's sky, while $b$ denotes the impact parameter. 
}%%
\label{fig:trajectory}
\end{figure}

Now we consider the photon trajectory radiating from a neutron star surface at $r=R$, which approaches to an observer at a distance $D \gg R$ from the star. The equation of motion for massless particle is derived from the Euler-Lagrange equation. Then, the angle position, $\psi(R)$, as shown in Fig. \ref{fig:trajectory} is given by
\begin{equation}
  \psi(R) = \int_R^\infty \frac{dr}{C}\left[\frac{1}{AB}\left(\frac{1}{b^2}-\frac{A}{C}\right)\right]^{-1/2}, \label{eq:psi}
\end{equation}
where $b$ denotes an impact parameter \cite{PFC1983}. The impact parameter is associated with the emission angle $\alpha$, which is the angle between the direction of photon emission and the normal vector at the emission point on the stellar surface, as
\begin{equation}
  \sin\alpha = b\sqrt{\frac{A(R)}{C(R)}}. \label{eq:b}
\end{equation}

%%%%%%%%%%%%%%%%%%%%%%%%%%%%%%%%%%%
% Figure 1
%%%%%%%%%%%%%%%%%%%%%%%%%%%%%%%%%%%
%%%%%%%%%%%%%%%%%%%%%%%%%%%%%%%%%%%

Using Eqs. (\ref{eq:psi}) and (\ref{eq:b}), one can numerically obtain the relation between $\psi(R)$ and $\alpha$ for given $R$. The value of $\psi$ increases with $\alpha$, and becomes the maximum value of $\psi_{\rm cri}$ when $\alpha$ is equal to $\pi$. That is, the photon from the stellar surface at $\psi=\psi_{\rm cri}$ is tangentially emitted and reaches the observer, while the photon from the stellar surface for $\psi > \psi_{\rm cri}$, which corresponds to $\alpha>\pi$, can not reach the observer anymore. So, the region for $\psi > \psi_{\rm cri}$ becomes an invisible zone. Since the light bending occurs due to the relativistic effect, one can expect that $\psi_{\rm cri}$ becomes larger (or the invisible zone becomes smaller) as the gravitational field becomes stronger, where the neutron star compactness is a kind of the parameter expressing how strong the gravitational field is. In fact, it is shown in Ref. \cite{SM2018a} that $\psi_{\rm cri}$ monotonically increases as the stellar compactness increases and one can not theoretically exclude the existence of highly compact neutron stars, where $\psi_{\rm cri}$ becomes more than $\pi$. If such a highly compact neutron star would exist, the invisible zone disappears and the photon emitted from any position on the neutron star surface can reach the observer, where the number of photon paths may be more than one, depending on the stellar compactness and the angle position of photon emission \cite{SM2018a}. In this article, we especially focus on such highly compact neutron stars, taking into account the finite size effect of hot spot, while the previous study has been done with the pointlike approximation of hot spot \cite{SM2018a}.

In order to calculate the photon flux measured by an observer, one has to integrate the flux emitted from the hot spot on the neutron star surface. In general, the bolometric intensity, $I_0$, measured in the vicinity of the stellar surface depends on the emission angle due to scattering in the neutron star atmosphere \cite{Zavlin:1996wd,2006ApJ...644.1090H,2012ApJ...749...52H,Salmi:2019pod}, but we simply assume the uniform emission, i.e., $I_0= {\rm const.}$, as in the previous studies (e.g., \cite{Beloborodov2002,SM2018b,SSP2019}). Then, the bolometric flux, $F$, radiated from the area of hot spot, $S$, on the stellar surface is calculated via
\begin{gather}
  F=F_1\iint_S \sin\alpha \cos\alpha\frac{d\alpha}{d\psi}d\psi d\phi, \label{eq:FF} \\
  S = C(R)\iint_S \sin\psi d\psi d\phi,  \label{eq:SS}
\end{gather}
where 
\begin{equation}
  F_1\equiv \frac{I_0A(R)C(R)}{D^2}   \label{eq:F1}
\end{equation}
and $\iint_S$ denotes that the integration should be done only inside the hot spot \cite{SSP2019}.

%%%%%%%%%%%%%%%%%%%%%%%%%%%%%%%%%%%%%%%%%%%%%%%%
\section{Light curves from slowly rotating neutron stars}
\label{sec:III}
%%%%%%%%%%%%%%%%%%%%%%%%%%%%%%%%%%%%%%%%%%%%%%%%

\begin{figure}[tbp]
\begin{center}
\includegraphics[scale=0.4]{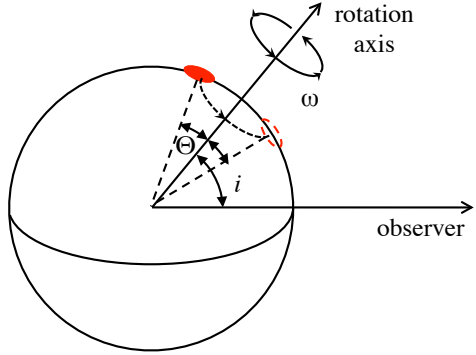}  
\end{center}
\caption{%%
Image of single hot spot on the slowly rotating neutron star characterized by two angles $\Theta$ and $i$.
}%%
\label{fig:pulsar}
\end{figure}

We particularly consider the light curve from a single hot spot on slowly rotating neutron stars in this study as in Ref. \cite{SSP2019}. That is, one can neglect the rotational effects, such as the Doppler, aberration, and time delay, and discuss the light curve in the Schwarzschild spacetime. We also focus on a circular hot spot with an opening angle $\Delta\psi$, an annular hot spot as shown in Fig. \ref{fig:pulsar3}, and a double circular hot spot as shown in Fig. \ref{fig:pulsar4},  although the light curves may generally depend on the shape of hot spot. In this study, we consider the rotating neutron stars with an angular velocity, $\omega$, by introducing two specific angles, $\Theta$ and $i$, as shown in Fig. \ref{fig:pulsar}, where $\Theta$ is the angle between the rotational axis and the normal vector at the center of hot spot, while $i$ is the angle between the rotational axis and the direction to the observer. By setting $t=0$ when the hot spot is closest to the observer, the angular position of the center of hot spot, $\psi_*$, is given as a function of time, $t$, as
\begin{equation}
  \cos\psi_* = \sin i \sin\Theta\cos (\omega t) + \cos i \cos\Theta. \label{eq:psi*}
\end{equation}
From this equation, one can see that the light curve for $(\Theta,i)=(\theta_1,\theta_2)$ is the same as that for $(\Theta,i)=(\theta_2,\theta_1)$. In addition, one can find the symmetry in the light curve, i.e., the amplitude at the time of $t$ for $T/2\le t\le T$ becomes the same as that at the time of $T-t$, using the rotational period $T$ defined by $T\equiv 2\pi/\omega$. 
%Thus, in this study we consider the light curves only for $0\le t\le T/2$.

%%%%%%%%%%%%%%%%%%%%%%%%%%%%%%%%%%%
% Figure 2
%%%%%%%%%%%%%%%%%%%%%%%%%%%%%%%%%%%
%%%%%%%%%%%%%%%%%%%%%%%%%%%%%%%%%%%

Owing to the nature of spherically symmetric spacetime, one can select the coordinate at any time so that the center of the hot spot is located at $(\psi,\phi)=(\psi_*,0)$ (as shown in Fig. 2 in Ref. \cite{SSP2019}). Then, the position inside the hot spot, $(\psi,\phi)$, should basically fulfill the condition as
\begin{equation}
  \cos(\Delta\psi) \le \sin\psi_* \sin\psi\cos\phi + \cos\psi_*\cos\psi. \label{eq:phi}
\end{equation}
With this condition, one can determine the boundary of hot spot in the $\phi$ direction for given $\psi$ as $\phi(\psi)$. So, from Eq. (\ref{eq:FF}) the bolometric flux from the circular hot spot, measured by an observer, is given by
\begin{equation}
  F(\psi_*,\Delta\psi) = 2F_1 \int_{\psi_*-\Delta\psi}^{\psi_*+\Delta\psi} \phi(\psi) \sin\alpha \cos\alpha \frac{d\alpha}{d\psi}d\psi,
   \label{eq:Fdpsi}
\end{equation}
for given values of $\psi_*$ and $\Delta\psi$. That is, once two angles $\Theta$ and $i$ and the opening angle $\Delta\psi$ are selected for the neutron star with $M$ and $R$, one can calculate the time variation of the bolometric flux with $\psi_*(t)$ determined from Eq. (\ref{eq:psi*}), which corresponds to the light curve. As in Ref. \cite{SSP2019}, we consider the light curves only for $\Theta=i$ in this study, where we especially focus on the case with $\Theta=i=30^{\circ}$, $60^{\circ}$, and $90^{\circ}$.

%%%%%%%%%%%%%%%%%%%%%%%%%%%%%%%%%%%%%%%%%%%%%%%%
\subsection{Dependence of the flux on $\Delta\psi$ and $M/R$}
\label{sec:IIIa}
%%%%%%%%%%%%%%%%%%%%%%%%%%%%%%%%%%%%%%%%%%%%%%%%

\begin{figure*}[tbp]
\begin{center}
\includegraphics[scale=0.36]{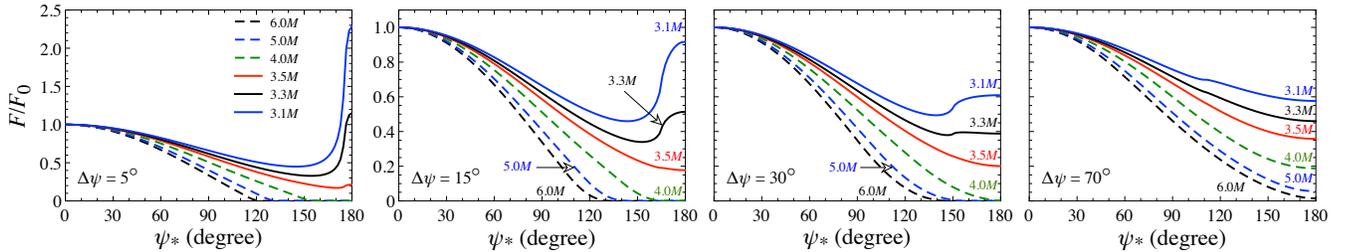}
\end{center}
\caption{%%
Normalized flux, $F/F_0$, is shown as a function of the angle of the center of hot spot, $\psi_*$, for the neutron star models with various compactness, i.e., $R=6.0M$, $5.0M$, $4.0M$, $3.5M$, $3.3M$, and $3.1M$, where $F_0$ denotes the flux with $\psi_*=0^{\circ}$. The panels from left to right correspond to the case with $\Delta\psi = 5^{\circ}$, $15^{\circ}$, $30^{\circ}$, and $70^{\circ}$. We note that the corresponding compactness is $M/R=0.167$, 0.200, 0.250, 0.286, 0.303, and 0.323.
}%%
\label{fig:F-psi}
\end{figure*}

Before considering the concrete light curves from the neutron stars, we see how the flux calculated with Eq. (\ref{eq:Fdpsi}) depends on the opening angle and the stellar compactness. In Fig. \ref{fig:F-psi}, the bolometric flux normalized by the flux with $\psi_*=0$, $F_0$, is shown as a function of $\psi_*$ for neutron star models with different compactness, where the panels from left to right correspond to the flux with $\Delta\psi = 5^{\circ}$, $15^{\circ}$, $30^{\circ}$, and $70^{\circ}$. In this figure, the dashed (solid) lines correspond to the stellar models whose compactness is less (larger) than 0.2840, i.e., the stellar models shown with the dashed (solid) lines have (do not have) an invisible zone at the backside of the star. Thus, if one adopts a pointlike approximation ($\Delta\psi \approx 0$), the flux from the neutron star becomes zero when the hot spot enters the invisible zone. But, once one considers the size effect, even if the center of hot spot enters the invisible zone, one could observe the non-zero flux from the neutron star during a part of the hot spot is still outside the invisible zone, as pointed out in Ref. \cite{SSP2019}. In practice, as shown in Fig. \ref{fig:F-psi}, one can observe that the region where the flux becomes zero decreases as the opening angle $\Delta \psi$ increases, and finally with $\Delta\psi=70^{\circ}$ one can observe the non-zero flux for the neutron star model with $R=6M$, even when $\psi_*=180^{\circ}$, i.e., the center of hot spot comes to the opposite side to the observer. That is, the area of hot spot with $\Delta\psi=70^{\circ}$ becomes larger than that of the invisible zone. 

 %%%%%%%%%%%%%%%%%%%%%%%%%%%%%%%%%%%
% Figure 3
%%%%%%%%%%%%%%%%%%%%%%%%%%%%%%%%%%%
%%%%%%%%%%%%%%%%%%%%%%%%%%%%%%%%%%%

%%%%%%%%%%%%%%%%%%%%%%%%%%%%%%%%%%%
% Figure 4
%%%%%%%%%%%%%%%%%%%%%%%%%%%%%%%%%%%
\begin{figure*}[tbp]
\begin{center}
\begin{tabular}{cc}
\includegraphics[scale=0.5]{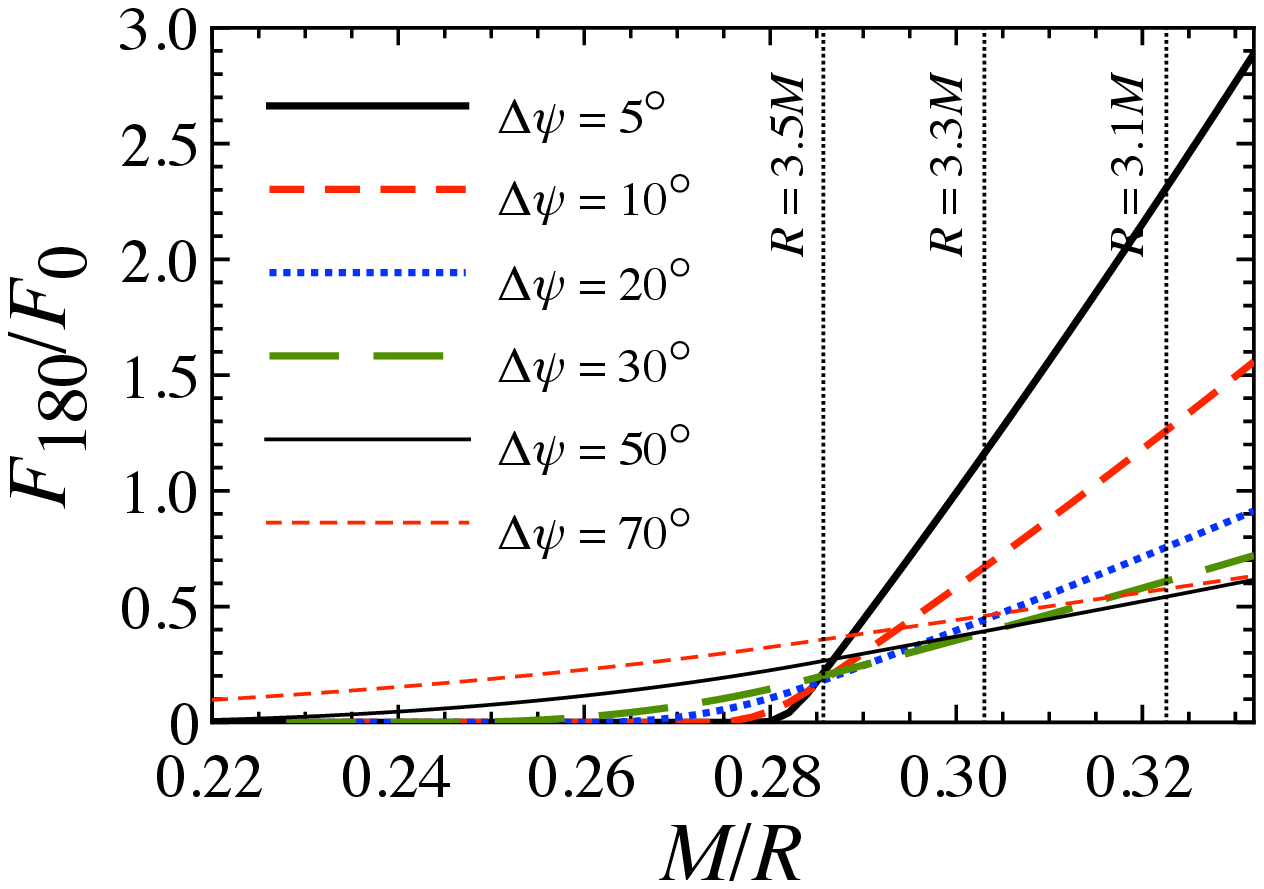} &
\includegraphics[scale=0.5]{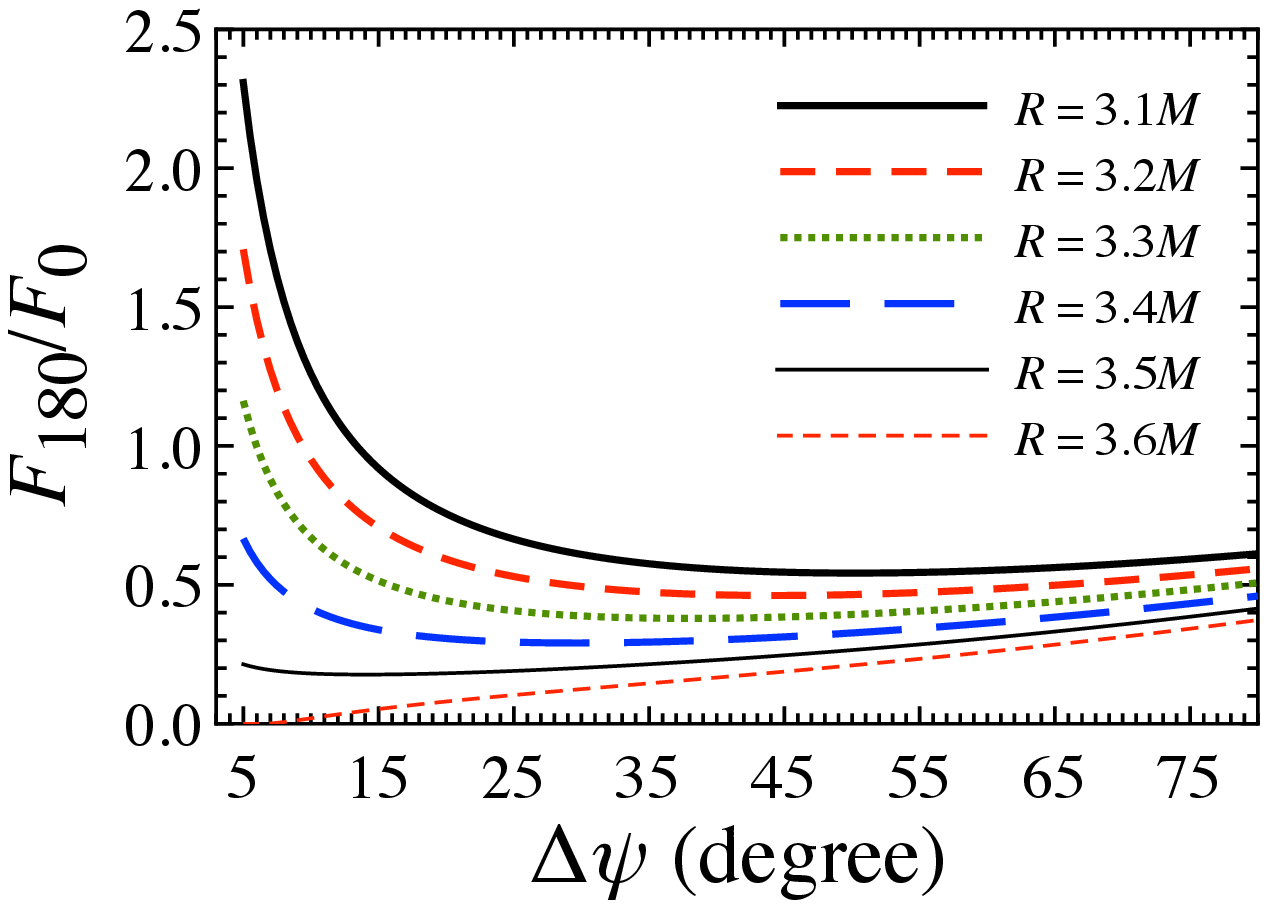}  
\end{tabular}
\end{center}
\caption{%%
Ratio of the flux with $\psi_*=180^{\circ}$ ($F_{180}$) to $F_{0}$ is shown as a function of the stellar compactness with fixing the opening angle ($\Delta\psi$) of hot spot in the left panel, while the same ratio is shown as a function of $\Delta\psi$ with fixing the stellar compactness in the right panel. 
}%%
\label{fig:FF}
\end{figure*}
%%%%%%%%%%%%%%%%%%%%%%%%%%%%%%%%%%%

On the other hand, in Fig. \ref{fig:F-psi} one also observes the brightening in the flux for the neutron star with larger compactness  when the hot spot comes to the backside of the star, which becomes significant as the hot spot area (or the opening angle) becomes small and as the stellar compactness increases. In order to clearly see this phenomenon, in Fig. \ref{fig:FF} we show the ratio of the flux with $\psi_*=180^{\circ}$ ($F_{180}$) to $F_0$ as a function of the stellar compactness with various opening angle in the left panel and as a function of the opening angle with various stellar compactness in the right panel. From the left panel, one can observe that the value of $F_{180}/F_0$ is almost proportional to $M/R$ for $M/R\gsim 0.285$, where the proportionality factor becomes larger as the opening angle decreases. Meanwhile, from the right panel, one can also find that the brightening strongly depends on the opening angle especially for small opening angle, and expect that the value of $F_{180}/F_0$ diverges in the limit of $\Delta\psi\to 0$ \cite{SM2018a}. Anyway, from Figs. \ref{fig:F-psi} and \ref{fig:FF}, for a circular hot spot, one can expect that the sharp brightening in flux can occur only for the neutron star models with larger compactness and smaller hot spot, when the hot spot comes to the opposite side to the observer. In addition, due to this brightening, one may observe a signature of the finite-size effect of hot spot in light curves from a quite highly compact star when the edge of hot spot comes to the opposite side to the observer, i.e., $\psi* \simeq \pi - \Delta \psi$, even if the area of hot spot is relatively large. In fact, in Fig. \ref{fig:F-psi} one can see such a signature for the neutron star model with $R=3.1M$ at $\psi_*\simeq 150^{\circ}$ and $110^{\circ}$ for $\Delta \psi=30^{\circ}$ and $70^{\circ}$, respectively.

%%%%%%%%%%%%%%%%%%%%%%%%%%%%%%%%%%%%%%%%%%%%%%%%
\subsection{Classification of the number of photon paths}
\label{sec:IIIb}
%%%%%%%%%%%%%%%%%%%%%%%%%%%%%%%%%%%%%%%%%%%%%%%%

%%%%%%%%%%%%%%%%%%%%%%%%%%%%%%%%%%%
% Figure 5
%%%%%%%%%%%%%%%%%%%%%%%%%%%%%%%%%%%
\begin{figure}[tbp]
\begin{center}
\begin{tabular}{c}
\includegraphics[scale=0.5]{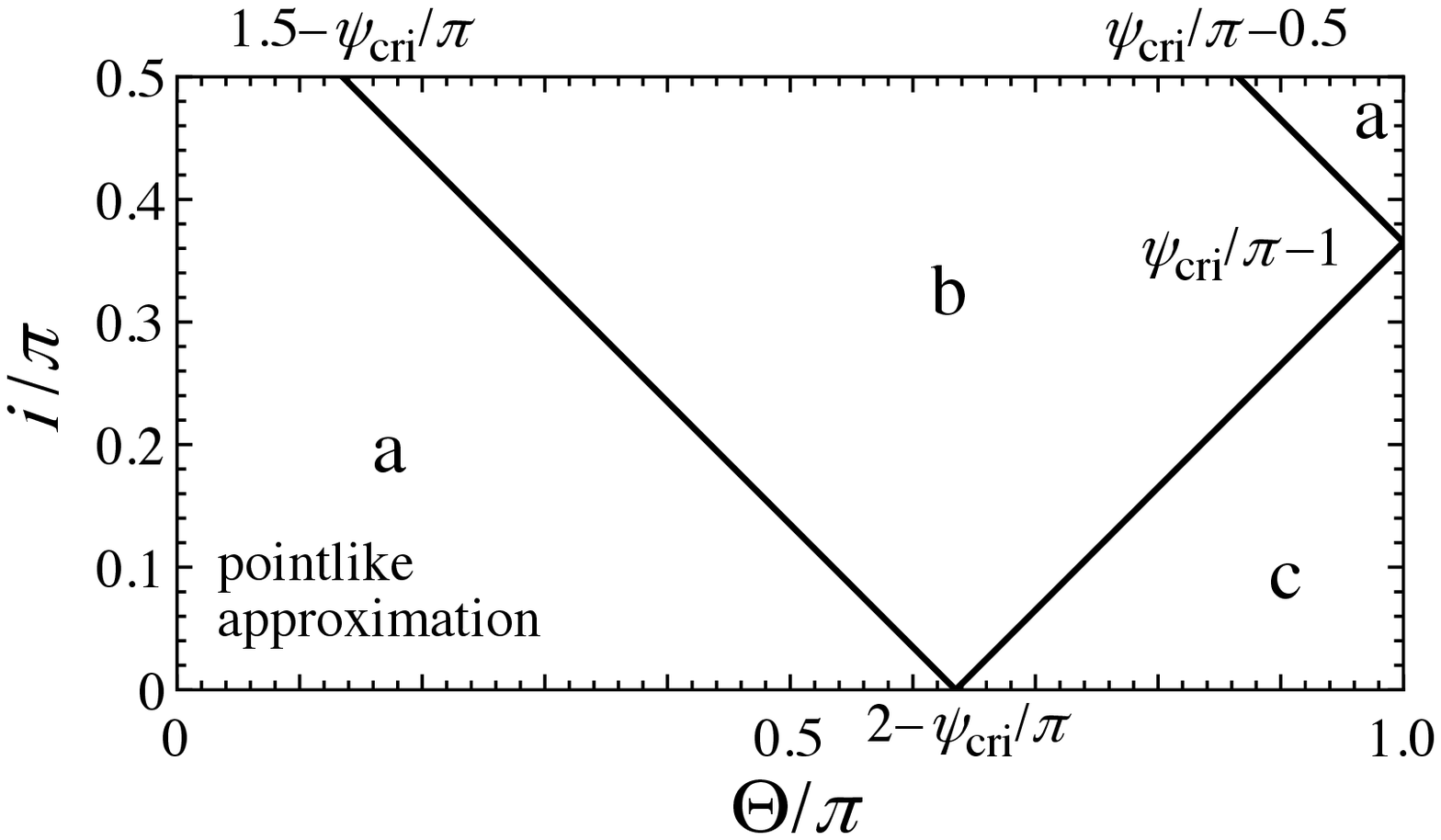} \\
\includegraphics[scale=0.5]{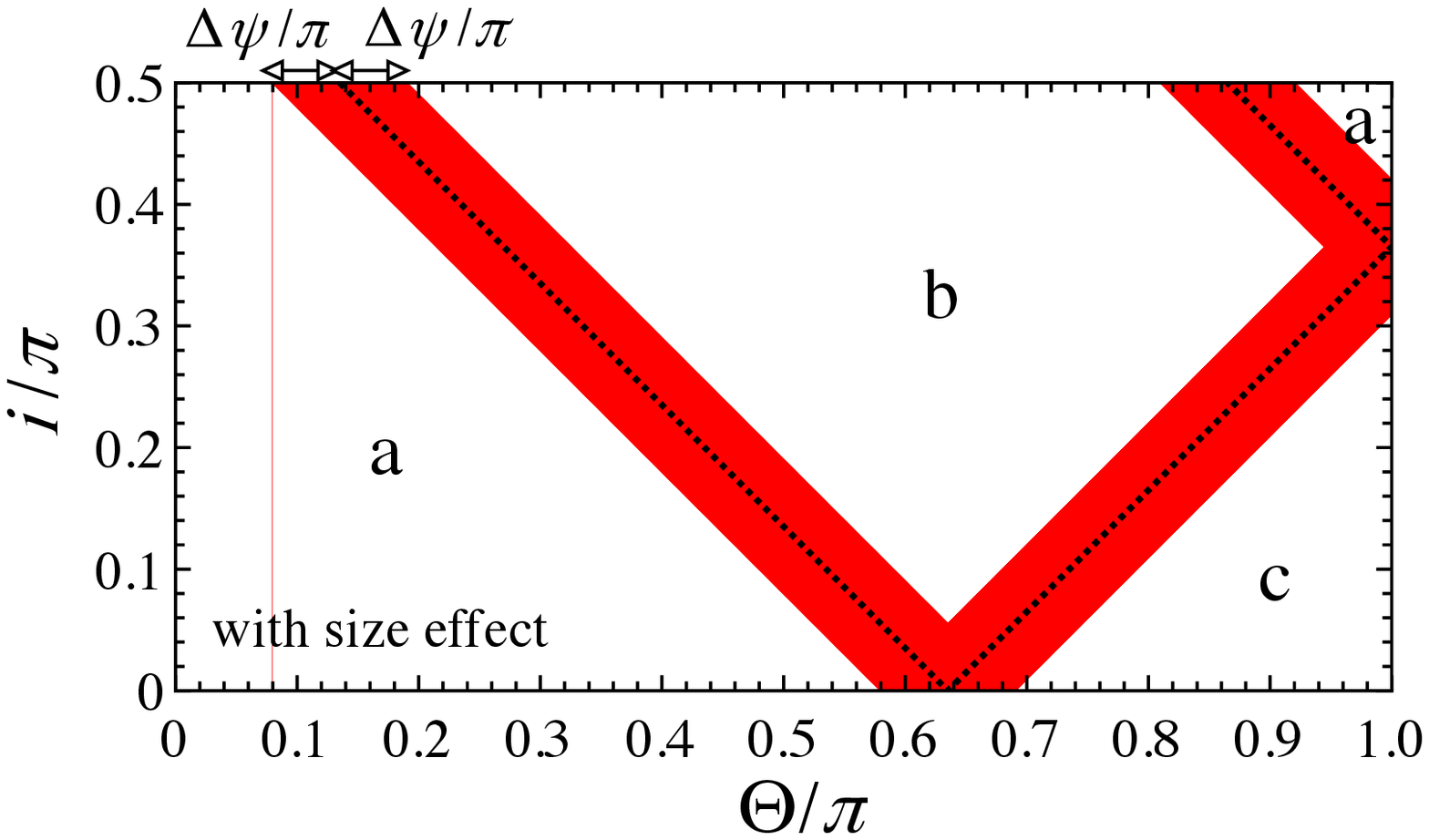} 
\end{tabular}
\end{center}
\caption{%%
Classification of the number of photon paths for the case of $\pi < \psi_{\rm cri} < 3\pi/2$. Depending on the value of $\psi_{\rm cri}$, three different situations can be considered as a function of $\Theta$ and $i$ (see text for details). The top panel corresponds to the case with pointlike approximation ($\Delta\psi\approx 0$), while the bottom panel corresponds to the case with the finite sizes effect of hot spot, where as an example $\Delta\psi$ is selected to be $\Delta\psi=10^{\circ}$.
}%%
\label{fig:class}
\end{figure}
%%%%%%%%%%%%%%%%%%%%%%%%%%%%%%%%%%%

As mentioned before, in this study we focus on the highly compact neutron stars, i.e., $M/R>0.2840$, where the invisible zone disappears. In this case, by definition the photon radiated from any position on the neutron star surface can reach an observer, while one has to count the number of photon paths depending on the angular position $\psi$ of photon emission \cite{SM2018a}. We remark that, as the value of $\psi_{\rm cri}$ increases (or as the stellar compactness increases), the maximum number of photon paths can also increase. Even so, in this study we consider the simplest case, such as $\pi < \psi_{\rm cri} < 3\pi/2$, which corresponds to $0.2840<M/R<0.3236$. Of course, one can consider the neutron star model with $M/R>0.3236$, but such a stellar model is very limited (see Fig. 3 in Ref. \cite{SM2018a}).

For the case of $\pi < \psi_{\rm cri} < 3\pi/2$, within the pointlike approximation, one can classify the situation as a function of $\Theta$ and $i$, as shown in the top panel of Fig. \ref{fig:class}, where the regions denoted by a, b, and c correspond to the followings.
\begin{enumerate}
  \item region a: number of photon paths is always only one,
  \item region b: number of photon paths becomes two for a fraction of the rotational period,
  \item region c: number of photon paths is always two.
\end{enumerate}
If one consider the finite size effect of the hot spot, the boundaries shown in the top panel of Fig. \ref{fig:class} become cloudy, as shown in the bottom panel of Fig. \ref{fig:class}, where the painted region corresponds to the situation that a part of the hot spot crosses over the boundary shown in the top panel of Fig. \ref{fig:class}. Anyway, if the number of photon paths becomes two, the total of two fluxes should be the observed flux.

%%%%%%%%%%%%%%%%%%%%%%%%%%%%%%%%%%%%%%%%%%%%%%%%
\subsection{Light curves with a circular hot spot}
\label{sec:IIIc}
%%%%%%%%%%%%%%%%%%%%%%%%%%%%%%%%%%%%%%%%%%%%%%%%

%%%%%%%%%%%%%%%%%%%%%%%%%%%%%%%%%%%
% Figure 6
%%%%%%%%%%%%%%%%%%%%%%%%%%%%%%%%%%%
\begin{figure*}[tbp]
\begin{center}
\begin{tabular}{cc}
\includegraphics[scale=0.5]{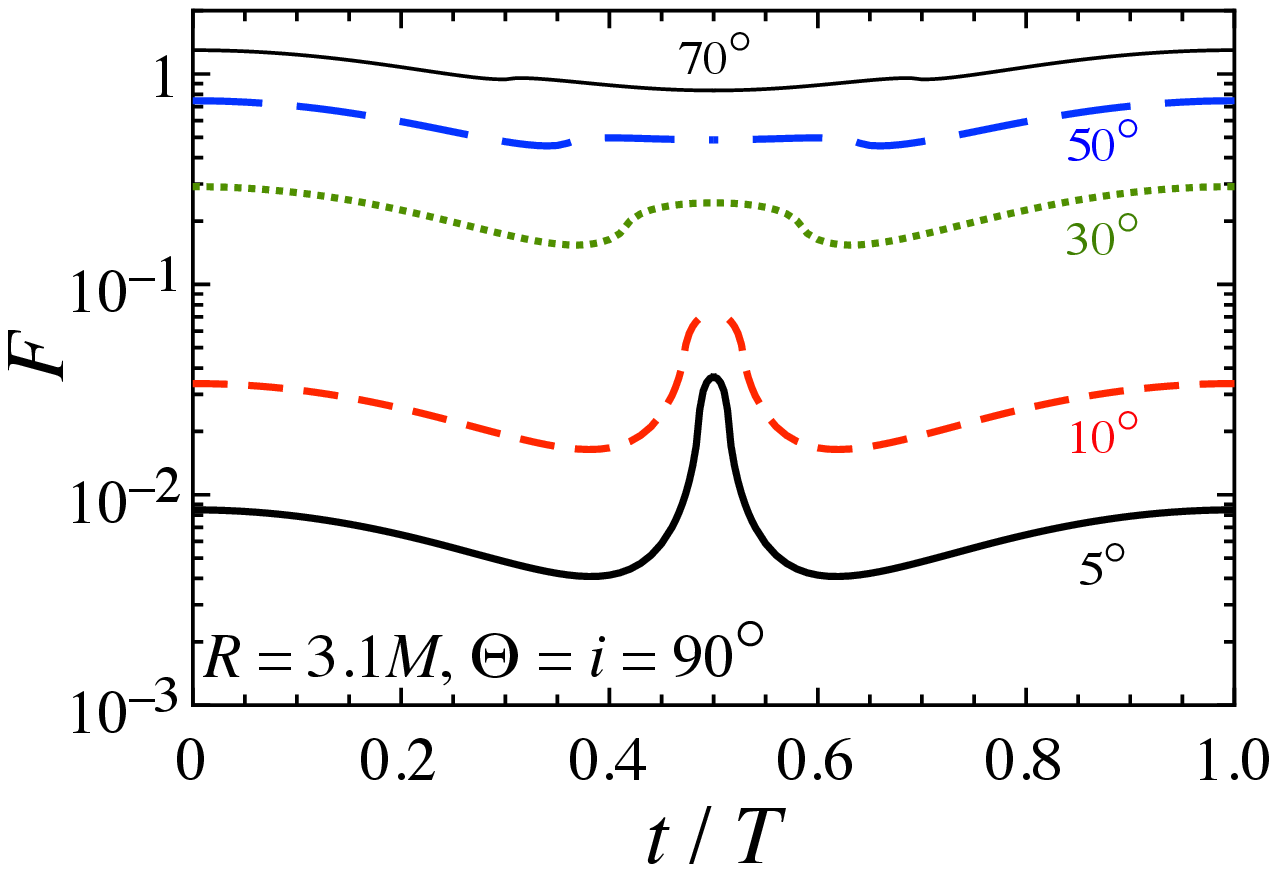} &
\includegraphics[scale=0.5]{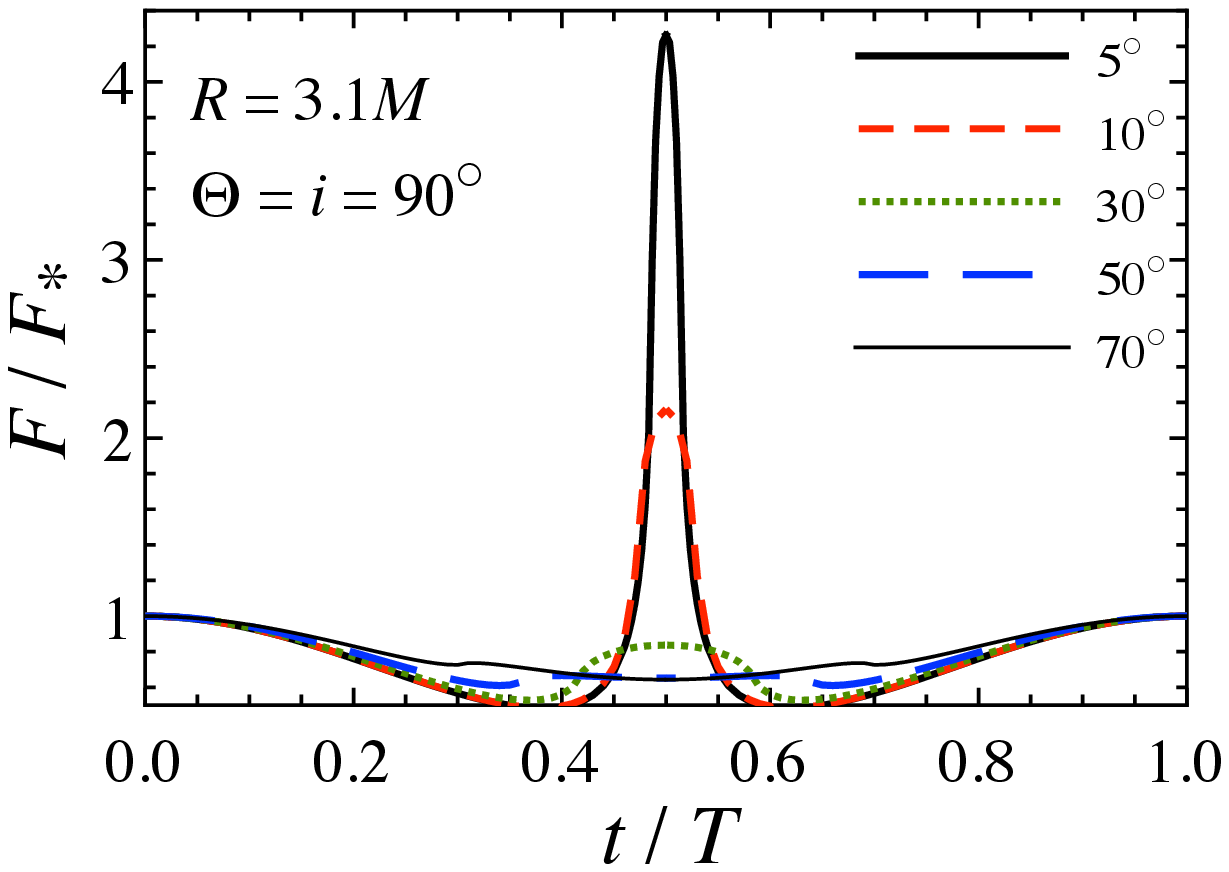} 
\end{tabular}
\end{center}
\caption{%%
Bolometric flux $F$ and $F$ normalized by the flux at $t=0$ ($F_*$) are respectively shown in the left and right panels for one rotational period for the neutron star model with $R=3.1M$ and $\Theta= i =90^{\circ}$, where the results with various opening angle $\Delta\psi$ are shown with different lines. 
}%%
\label{fig:Fct-31M90}
\end{figure*}
%%%%%%%%%%%%%%%%%%%%%%%%%%%%%%%%%%%

%%%%%%%%%%%%%%%%%%%%%%%%%%%%%%%%%%%
% Figure 7
%%%%%%%%%%%%%%%%%%%%%%%%%%%%%%%%%%%
\begin{figure*}[tbp]
\begin{center}
\includegraphics[scale=0.47]{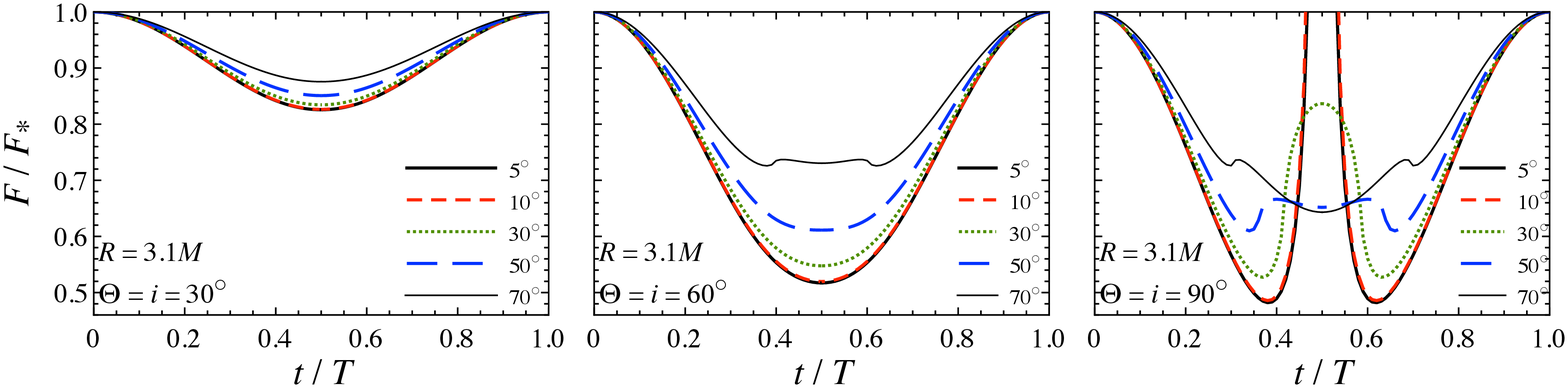}
\end{center}
\caption{%%
Same as the right panel of Fig. \ref{fig:Fct-31M90}, but for $\Theta= i =30^{\circ}$ (left), $60^{\circ}$ (middle), and $90^{\circ}$ (right). That is, the right panel is the enlarged view of the right panel of Fig. \ref{fig:Fct-31M90}.
}%%
\label{fig:Fct-31M306090}
\end{figure*}
%%%%%%%%%%%%%%%%%%%%%%%%%%%%%%%%%%%

\begin{figure*}[tbp]
\begin{center}
\includegraphics[scale=0.47]{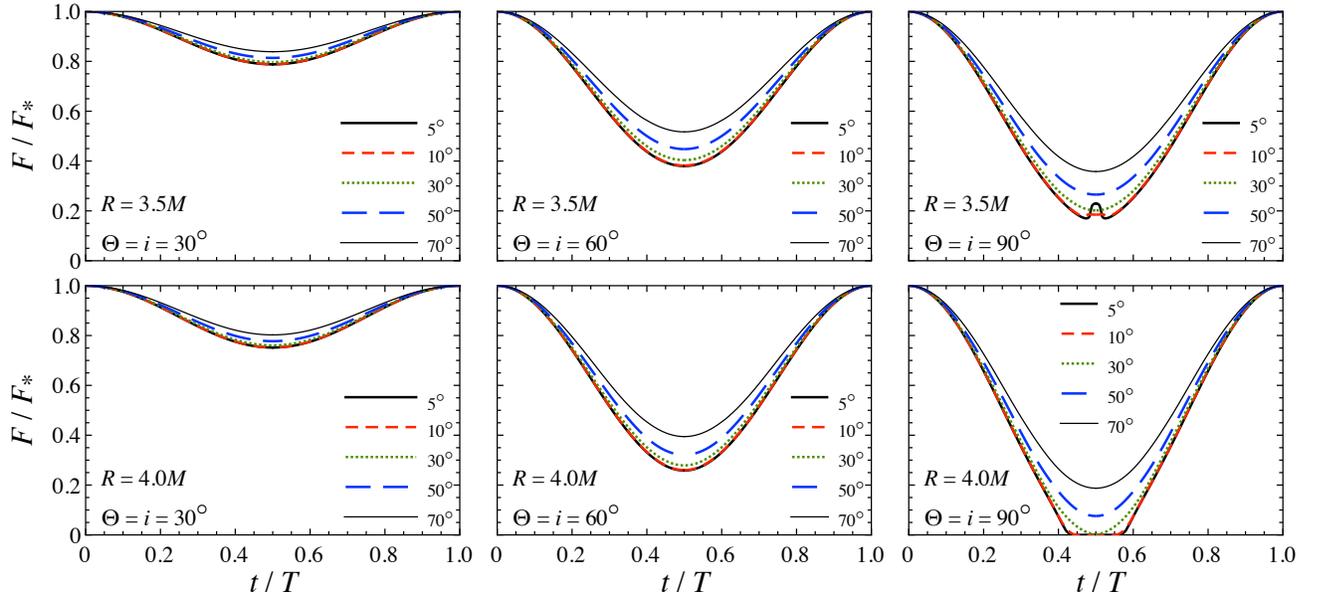} 
\end{center}
\caption{%%
Same as Fig. \ref{fig:Fct-31M306090}, but for the neutron star model with $R=3.5M$ (upper panels) and with $R=4.0M$ (lower panels). 
}%%
\label{fig:Fct-3540M306090}
\end{figure*}

Now, we consider the light curves from a slowly rotating neutron star with a circular hot spot. For considering the concrete light curves, we especially focus on the neutron star models with $R=3.1M$ and $3.5M$, where the corresponding values of $\psi_{\rm cri}$ are $\psi_{\rm cri} = 1.469\pi$ ($264.4^{\circ}$) and $1.010\pi$ ($181.8^{\circ}$), respectively. In addition, for reference we may  consider the neutron star model with $R=4.0M$, where $\psi_{\rm cri} = 0.848\pi$ ($152.6^{\circ}$), i.e., this model has an invisible zone. In Fig. \ref{fig:Fct-31M90}, we show the light curves for the neutron star model with $R=3.1M$ and $\Theta=i=90^{\circ}$, as varying the opening angle $\Delta\psi$ of hot spot, where the left and right panels correspond to the bolometric flux $F$ and $F$ normalized by the flux at $t=0$ ($F_*$). Since the hot spot with larger opening angle has lager area, the bolometric flux also becomes stronger as $\Delta\psi$ increases. Even so, one can obviously observe the brightening of flux for the hot spot with relatively small opening angle, when the hot spot comes to the opposite side to the observer (at $t=0.5T$), and the brightening effect becomes stronger as $\Delta\psi$ decreases. In the similar way, in Fig. \ref{fig:Fct-31M306090} we show $F/F_*$ for the neutron star model with $R=3.1M$ for $\Theta=i=30^{\circ}$ (left), $60^{\circ}$ (middle), and $90^{\circ}$ (right). That is, the right panel is just the enlarged view of the right panel of Fig. \ref{fig:Fct-31M90}. From this figure, one can see a simple behavior of the light curve even for highly compact neutron stars, if the hot spot can not reach the opposite side to the observer. On the other hand, with larger opening angle, one can observe a signature in the light curve when the edge of hot spot comes to the opposite side of the observer, where the condition given by 
\begin{equation}
  \Delta\psi + \psi_* = \pi \label{eq:cond}
\end{equation}
should be satisfied at that time. Since $\psi_*$ is determined via Eq. (\ref{eq:psi*}), one may constrain the combination of the angles of $\Theta$, $i$, and $\Delta\psi$ with the observation of light curve. That is, with Eqs. (\ref{eq:psi*}) and (\ref{eq:cond}) one can derive the constraint as
\begin{equation}
  \frac{t}{T} \approx \arccos \left[\frac{\cos(\pi-\Delta\psi)-\cos i\cos\Theta}{2\pi\sin i\sin\Theta}\right]. \label{eq:cond'}
\end{equation}
In practice, for example, one can derive $t/T=0.395$ (and $t/T=1-0.395=0.605$) for the hot spot with $\Theta=i=60^{\circ}$ and $\Delta\psi=70^{\circ}$ via Eq. (\ref{eq:cond'}), while at that time one can clearly observe a signature in the light curve as shown in the middle panel of Fig. \ref{fig:Fct-31M306090}. Furthermore, in Fig. \ref{fig:Fct-3540M306090} we show the light curve from neutron star models with lower compactness, where the top and bottom panels correspond to the light curves for the neutron star models with $R=3.5M$ and $4.0M$. The neutron star model with $R=3.5M$ does not have an invisible zone as mentioned before, but the relativistic effect is not so strong that the brightening in light curve is very weak. This a reason why one can see tiny brightening only for the hot spot with $\Delta\psi=5^{\circ}$ for $\Theta=i=90^{\circ}$ and why one can not observe a signature when the edge of the hot spot with relatively large opening angle comes to the opposite side to the observer.  Additionally, one can see that the light curves with smaller value of $\Theta$ and $i$, e.g., $\Theta=i=30^{\circ}$, are almost independent of the stellar compactness.

%%%%%%%%%%%%%%%%%%%%%%%%%%%%%%%%%%%
% Figure 8
%%%%%%%%%%%%%%%%%%%%%%%%%%%%%%%%%%%
%%%%%%%%%%%%%%%%%%%%%%%%%%%%%%%%%%%

%%%%%%%%%%%%%%%%%%%%%%%%%%%%%%%%%%%%%%%%%%%%%%%%
\subsection{Light curves with an annular hot spot}
\label{sec:IIId}
%%%%%%%%%%%%%%%%%%%%%%%%%%%%%%%%%%%%%%%%%%%%%%%%

\begin{figure}[tbp]
\begin{center}
\includegraphics[scale=0.4]{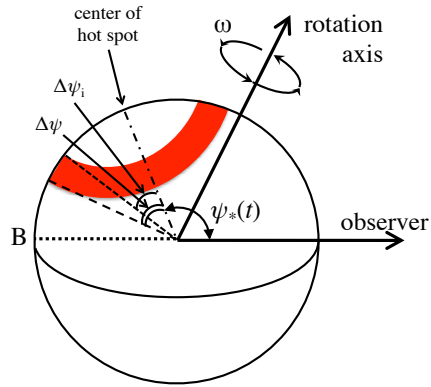}  
\end{center}
\caption{%%
Image of an annular geometry hot spot on the slowly rotating neutron star, where the width of hot spot is $\Delta\psi - \Delta \psi_i$ and the center of hot spot is $\psi_*$ determined from Eq. (\ref{eq:psi*}) at each time. The point B denotes the position on the neutron star opposite to the observer.
}%%
\label{fig:pulsar3}
\end{figure}

\begin{figure*}[tbp]
\begin{center}
\includegraphics[scale=0.47]{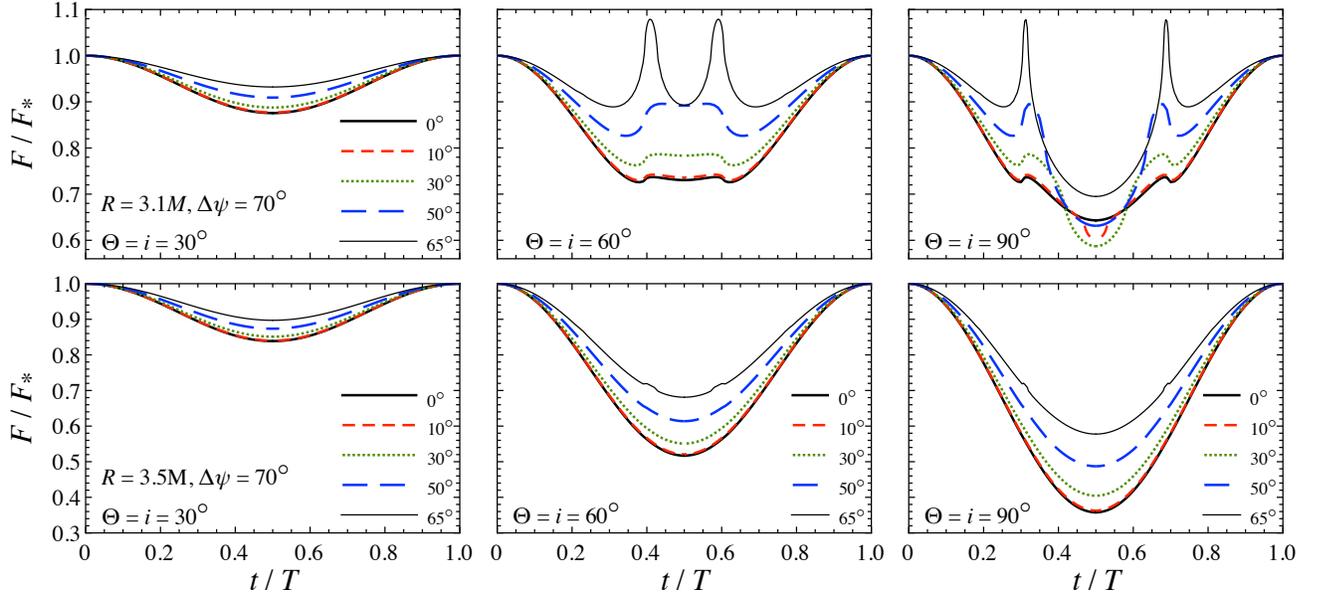} 
\end{center}
\caption{%%
Light curves from an annular hot spot with $\Delta\psi=70^{\circ}$. The top and bottom panels correspond to the light curves from the neutron star model with $R=3.1M$ and $3.5M$, while the left, middle, and right panels correspond to the case with $\Theta=i=30^{\circ}$, $60^{\circ}$, and $90^{\circ}$. In each panel, the different lines denote the results with different values of $\Delta\psi_i$.
}%%
\label{fig:Fat-dp70}
\end{figure*}

Next, we consider the light curves from the slowly rotating neutron star with an annular hot spot, as shown in Fig. \ref{fig:pulsar3}. In this case, the bolometric flux can be calculated as
\begin{equation}
  F_{\rm an} = F(\psi_*,\Delta\psi) - F(\psi_*,\Delta\psi_i), \label{eq:Fan}
\end{equation}
using the flux for a circular hot spot given by Eq. (\ref{eq:Fdpsi}). The light curves with $\Delta\psi=70^{\circ}$ are shown in Fig. \ref{fig:Fat-dp70}, as changing the value of $\Delta\psi_i$, where the panels from left to right correspond to the results for $\Theta=i=30^{\circ}$, $60^{\circ}$, and $90^{\circ}$, while top and bottom panels correspond to the results for the neutron star model with $R=3.1M$ and $3.5M$. The case with $\Delta\psi_i=0^{\circ}$ is the same as the light curves from a circular hot spot with $\Delta\psi=70^{\circ}$. From this figure, one can see a signal in the light curves when the hot spot crosses the opposite side to the observer, i.e., the point B in Fig. \ref{fig:pulsar3}. In particular, as the width of annular hot spot is narrower, such signals are more significant. In addition, even for the neutron star model with $R=3.5M$, one can see a tiny signature with $\Delta\psi_i=65^{\circ}$. We remark that, since this signature corresponds to the time when the band of a annular hot spot comes to the opposite side to the observer, the corresponding time should be determined with only geometry, such as the angles of $\Theta$, $i$, and $\Delta\psi$. That is, such a time is almost independent of the stellar compactness, i.e., $t/T\simeq 0.41$ (and  $0.59$) for $\Theta=i=60^{\circ}$, while $t/T\simeq 0.31$ (and $0.69$) for $\Theta=i=90^{\circ}$. Thus, by observing the light curve one may constrain the combination of the angles of $\Theta$, $i$, and $\Delta\psi$ again, which is similar to the discussion with a circular hot spot. Anyway, since these signatures appear when the spot region comes to the opposite side to the observer, one could constrain the geometry as
\begin{equation}
   \Theta + i + \Delta\psi \ge \pi, \label{eq:cond1}
\end{equation}
if such signatures would be observed. On the other hand, as in the case with a circular hot spot, since the hot spot does not reach the opposite side to the observer for the neutron star model with $\Theta=i=30^{\circ}$, one observes a simple behavior of light curves.

%%%%%%%%%%%%%%%%%%%%%%%%%%%%%%%%%%%
% Figure 9
%%%%%%%%%%%%%%%%%%%%%%%%%%%%%%%%%%%
%%%%%%%%%%%%%%%%%%%%%%%%%%%%%%%%%%%

%%%%%%%%%%%%%%%%%%%%%%%%%%%%%%%%%%%
% Figure 10
%%%%%%%%%%%%%%%%%%%%%%%%%%%%%%%%%%%
%%%%%%%%%%%%%%%%%%%%%%%%%%%%%%%%%%%

\begin{figure}[tbp]
\begin{center}
\begin{tabular}{c}
\includegraphics[scale=0.5]{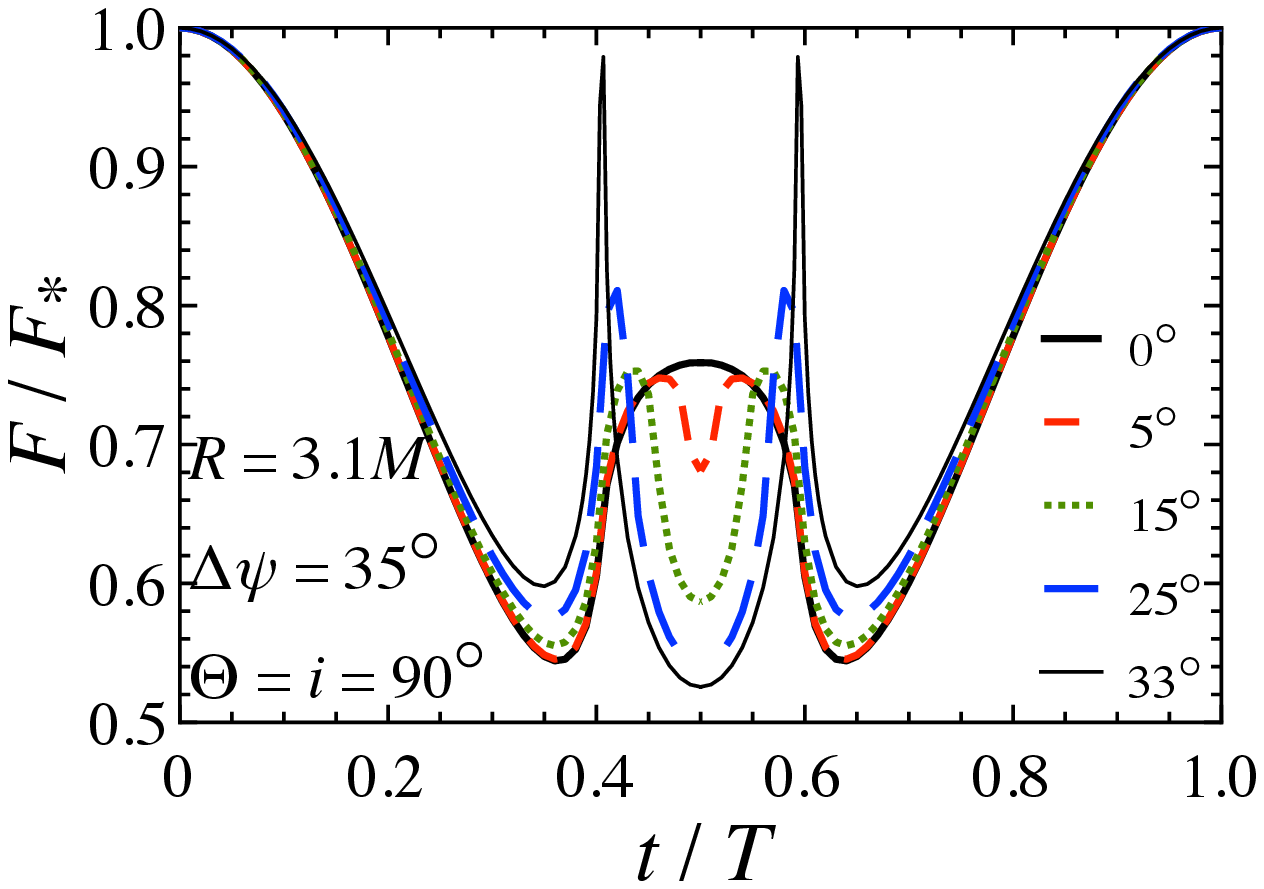} \\
\includegraphics[scale=0.5]{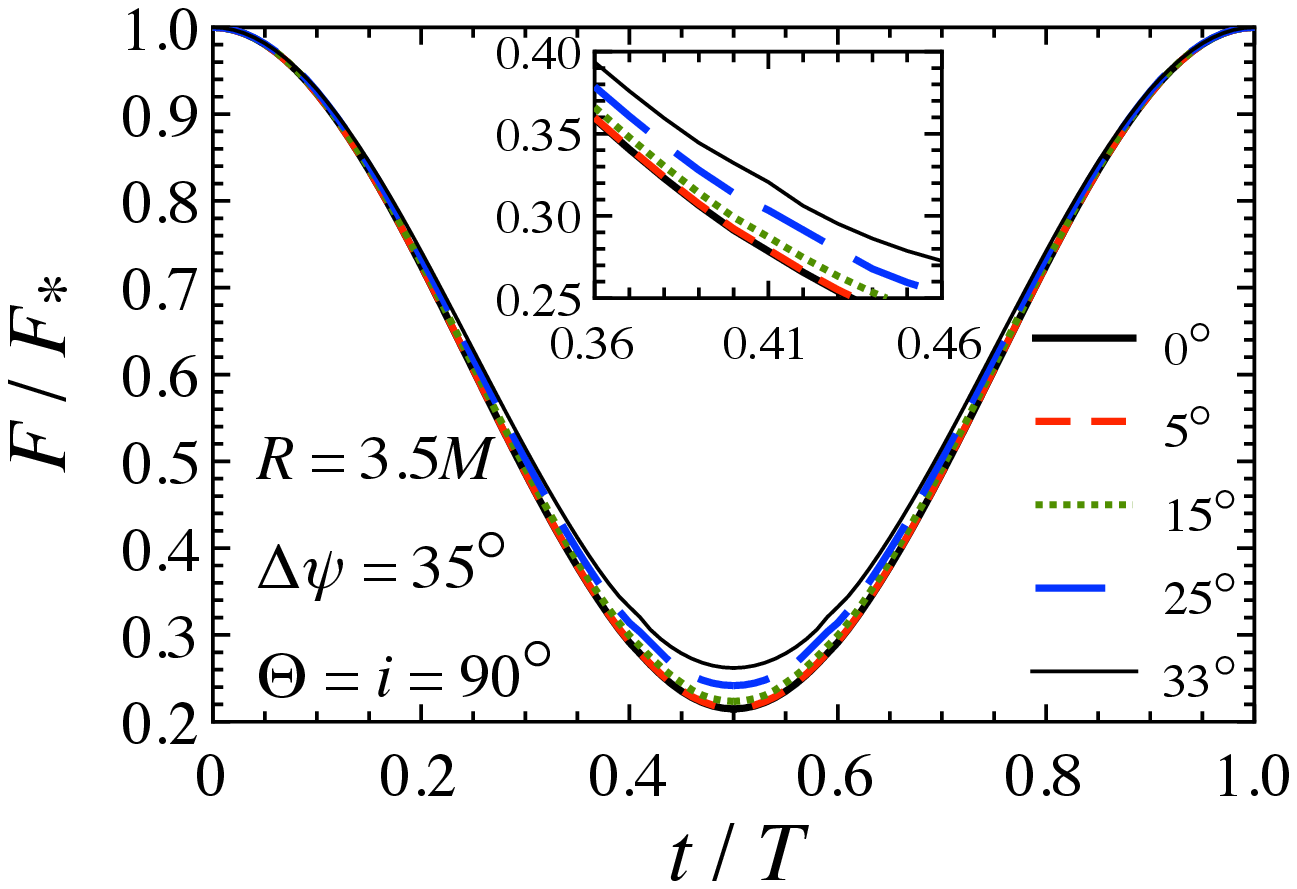} 
\end{tabular}
\end{center}
\caption{%%
Light curves from an annular hot spot with $\Delta\psi=35^{\circ}$ on the neutron star with $R=3.1M$ (top panel) and $3.5M$ (bottom panel) for $\Theta=i=90^{\circ}$. As in Fig. \ref{fig:Fat-dp70}, the different lines correspond to the results with different values of $\Delta\psi_i$.
}%%
\label{fig:Fat-dp35}
\end{figure}

In the same way, in Fig. \ref{fig:Fat-dp35} we show the light curves with $\Delta\psi=35^{\circ}$ and $\Theta=i=90^{\circ}$, as changing the value of $\Delta\psi_i$, where top and bottom panels correspond to the results for the neutron star model with $R=3.1M$ and $3.5M$. In this case with $\Delta\psi=35^{\circ}$, since the hot spot can not reach the opposite side to the observer for $\Theta=i=30^{\circ}$ and $60^{\circ}$, the corresponding light curve becomes very simple. On the other hand, for $\Theta=i=90^{\circ}$ one can observe the similar behavior as shown in Fig.\ref{fig:Fat-dp70}. That is, as the width of annular hot spot decreases, the signature when the band of annular hot spot comes to the opposite side to the observer becomes more significant, where the corresponding time depends on the angles of $\Theta$, $i$, and $\Delta\psi$, independently of the stellar compactness. So, even though the signature for the neutron star with $R=3.5M$ and $\Delta\psi_i=33^{\circ}$ is quite faint, one can see the signal when the band of hot spot crosses the opposite side to the observer at $t/T\simeq 0.41$.

%%%%%%%%%%%%%%%%%%%%%%%%%%%%%%%%%%%
% Figure 11
%%%%%%%%%%%%%%%%%%%%%%%%%%%%%%%%%%%
%%%%%%%%%%%%%%%%%%%%%%%%%%%%%%%%%%%

%%%%%%%%%%%%%%%%%%%%%%%%%%%%%%%%%%%%%%%%%%%%%%%%
\subsection{Light curves with a double circular hot spot}
\label{sec:IIId}
%%%%%%%%%%%%%%%%%%%%%%%%%%%%%%%%%%%%%%%%%%%%%%%%

\begin{figure}[tbp]
\begin{center}
\includegraphics[scale=0.4]{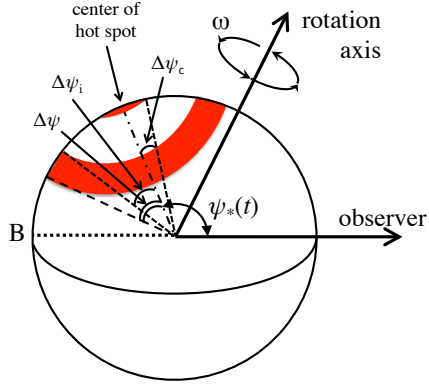}  
\end{center}
\caption{%%
Image of a double circular hot spot on the slowly rotating neutron star, where the width of outer annular hot spot is $\Delta\psi - \Delta \psi_i$ and the opening angle of the inner circular hot spot is $\Delta\psi_c$. The center of hot spot is $\psi_*$ determined from Eq. (\ref{eq:psi*}) at each time. 
}%%
\label{fig:pulsar4}
\end{figure}

\begin{figure}[tbp]
\begin{center}
\includegraphics[scale=0.5]{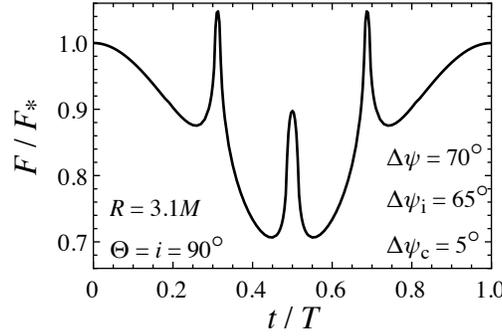}  
\end{center}
\caption{%%
Light curves from a double circular hot spot with $\Delta \psi=70^{\circ}$, $\Delta\psi_i=65^{\circ}$, and $\Delta \psi_c=5^{\circ}$ on the neutron star model with $R=3.1M$ for $\Theta=i=90^{\circ}$. 
}%%
\label{fig:Fdt-31MII90d70i65c05}
\end{figure}

Finally, we also consider the light curves from the slowly rotating neutron star with a double circular hot spot, as shown in Fig. \ref{fig:pulsar4}, where the bolometric flux can be calculated via
\begin{equation}
  F_{\rm db} = F(\psi_*,\Delta\psi) - F(\psi_*,\Delta\psi_i) + F(\psi_*,\Delta\psi_c). \label{eq:Fdb}
\end{equation}
As an example, we calculate light curves from the neutron star model with $R=3.1M$ and $\Theta=i=90^{\circ}$ espcially with $\Delta \psi=70^{\circ}$, $\Delta\psi_i=65^{\circ}$, and $\Delta \psi_c=5^{\circ}$. The resultant light curves are shown in Fig. \ref{fig:Fdt-31MII90d70i65c05}. From this figure, one can observe three peaks at $t/T\simeq 0.31$, $0.50$, and $0.69$. Obviously, the peaks at  $t/T\simeq 0.31$ and $0.69$ correspond to the time when the outer band of double circular spot crosses the opposite side to the observer (as same as in the top-right panel of Fig. \ref{fig:Fat-dp70}), while the peak at $t/T=0.50$ corresponds to the time when the central hot spot crosses the opposite side to the observer. That is, by observing the light curve from a highly compact neutron star,  one may know the number of narrow band region and/or the spot with tiny area even in a more complicated shape of hot spot, which could appear as peaks in the light curves when they cross the opposite side to the observer.

%%%%%%%%%%%%%%%%%%%%%%%%%%%%%%%%%%%
% Figure 12
%%%%%%%%%%%%%%%%%%%%%%%%%%%%%%%%%%%
%%%%%%%%%%%%%%%%%%%%%%%%%%%%%%%%%%%

%%%%%%%%%%%%%%%%%%%%%%%%%%%%%%%%%%%
% Figure 13
%%%%%%%%%%%%%%%%%%%%%%%%%%%%%%%%%%%
%%%%%%%%%%%%%%%%%%%%%%%%%%%%%%%%%%%

%%%%%%%%%%%%%%%%%%%%%%%%%%%%%%%%%%%%%%%%%%%%%%%%
\section{Conclusion}
\label{sec:IV}
%%%%%%%%%%%%%%%%%%%%%%%%%%%%%%%%%%%%%%%%%%%%%%%%

Light bending is one of the most important relativistic effects, while a property expressing the strength of a gravitational field around/inside a neutron star is the ratio of the mass to the radius, i.e., the compactness. So, since the light curve produced by a rotating neutron star strongly depends on its compactness, as an inverse problem one would extract the compactness of neutron star by carefully observing the light curve. In this study, we systematically examine the light curve from a single hot spot on a slowly rotating neutron star, especially focusing on a highly compact neutron star, where the so-called invisible zone disappears. In particular, we take into account the finite-size effect of the hot spot, where we adopt three different shapes of hot spot, i.e., a circular, annular, and double circular hot spot. As a result, we find that the light curve with any shapes of hot spot becomes very simple even for a highly compact neutrons star, if any position of the hot spot can not reach the opposite side to the observer. On the other hand, we find that the brightening in flux occurs when a hot spot with tiny area crosses the opposite side to the observer. So, via the direct observation of a light curve from a highly compact neutron star, one may constrain the geometry of neutron star, i.e., the combination of the inclination angle, the angle between the rotational axis and normal vector at the center of hot spot, and the opening angle of circular hot spot. In addition, we point out the possibility for extracting the number of the narrow band region and/or the spot with tiny area even in a more complicated shape of hot spot, which would appear as peaks in the light curve when such a tiny region crosses the opposite side to the observer. Furthermore, since the amplitude of these peaks depends on the stellar compactness, through the observation of light curve one may also constrain the stellar compactness.

%\newpage
%%%%%%%%%%%%%%%%%%%%%%%%%%%%%%%%%%%%%%%%%%%%%%%%
\acknowledgments
%%%%%%%%%%%%%%%%%%%%%%%%%%%%%%%%%%%%%%%%%%%%%%%%
%HS is grateful to T. Kawashima for giving valuable comments. 
This work was supported in part by Grant-in-Aid for Scientific Research (C) through Grant No. JP17K05458 provided by JSPS. 
%We are grateful to RIKEN Interdisciplinary Theoretical \& Mathematical Science Program (iTHEMS) for providing a facility. 

%\appendix
%%%%%%%%%%%%%%%%%%%%%%%%%%%%%%%%%%%%%%%%%%%%%%%%
%\section{Similarity of the light curve}
%\label{sec:a1}
%%%%%%%%%%%%%%%%%%%%%%%%%%%%%%%%%%%%%%%%%%%%%%%%

%%%%%%%%%%%%%%%%%%%%%%%%%%%%%%%%%%%%%%%%%%%%%%%%


\begin{thebibliography}{999}
\bibitem{example}
  We remark that the compactness of neutron star model with the minimum radius inferred from the GW170817 event,
  i.e., $R=10.68$ km with $M=1.6M_\odot$ \cite{Bauswein17}, is 0.2212, which is lower than the critical compactness $M/R=0.2840$ 
  because the mass is still small.
\end{thebibliography}
\end{document}